\documentclass[aps, prl, twocolumn, Letterpaper, nofootinbib, superscriptaddress, balancelastpage, preprintnumbers,longbibliography]{revtex4-1}

\usepackage{array}[=2016-10-06]
\usepackage{graphicx}
\usepackage{dcolumn}
\usepackage{amsmath, amsthm, mathrsfs, amsfonts, dsfont}
\usepackage{url}
\usepackage{bm, bbold, slashed}
\usepackage[dvipsnames]{xcolor}
\usepackage{easy-todo}
\usepackage{stmaryrd}
\usepackage[colorlinks,citecolor=blue,urlcolor=blue,linkcolor=blue]{hyperref}
\usepackage{braket}
\usepackage{subfig}
\usepackage{pifont}
\usepackage{booktabs}    
\usepackage{subcaption}  

\newcommand{\DB}{\scalebox{0.6}{\ensuremath{\mathrm{DB}}}}

\def\nn{\nonumber}
\newcommand{\cO}{\mathcal{O}}
\newcommand{\cQ}{\mathcal{Q}}

\newcommand\tickg{{\color{ForestGreen}\checkmark}}
\newcommand\crossr{{\color{RedOrange}\text{\sffamily X}}}

\newcolumntype{P}[1]{>{\centering\arraybackslash}p{#1}} 

\begin{document}

\preprint{QMUL-PH-25-19}

\title{Unexpected Symmetries of Kerr Black Hole Scattering}

\author{Dogan Akpinar}
\email{dogan.akpinar@ed.ac.uk} 
\affiliation{Higgs Centre for Theoretical Physics, School of Physics and
Astronomy, University of Edinburgh, EH9 3FD, UK} 
\author{Graham R. Brown}
\email{graham.brown@ed.ac.uk} 
\affiliation{Higgs Centre for Theoretical Physics, School of Physics and
Astronomy, University of Edinburgh, EH9 3FD, UK} 
\author{Riccardo Gonzo}
\email{r.gonzo@qmul.ac.uk} 
\affiliation{Centre for Theoretical Physics, Department of Physics and Astronomy,
Queen Mary University of London, Mile End Road, London E1 4NS, United Kingdom} 
\author{Mao Zeng}
\email{mao.zeng@ed.ac.uk} 
\affiliation{Higgs Centre for Theoretical Physics, School of Physics and
Astronomy, University of Edinburgh, EH9 3FD, UK}%

\date{\today}

\begin{abstract}
  Motivated by the recent introduction of the Dirac bracket framework to compute spinning observables for the scattering of Kerr black holes, we initiate the study of conserved quantities from an on-shell amplitude perspective. We establish new results for the conservation of energy, angular momentum, the R\"udiger invariant and the quadrupolar Carter constant using the spinning radial action extracted from the literature both in the probe limit and beyond, up to third post-Minkowskian order in the conservative sector. Furthermore, we offer a new perspective on the spin-shift symmetry of the radial action, clarifying its role in the dynamics. Finally, we define a new on-shell notion of asymptotic integrability in the Liouville sense and present strong evidence that it is surprisingly satisfied by a spinning probe in Kerr up to quartic order in the probe spin, to all orders in the post-Minkowskian expansion. We further establish integrability beyond the probe limit at low post-Minkowskian orders. Our results suggest important new implications for the dynamics of Kerr black holes.  
\end{abstract}

\maketitle

\section{Motivation and introduction}\label{sec:intro}

The groundbreaking detection of gravitational waves by the LIGO-Virgo-KAGRA collaboration~\cite{LIGOScientific:2016aoc,LIGOScientific:2017vwq} has opened new avenues for exploring the universe, enabling high-precision investigations of compact astrophysical objects such as black holes and neutron stars. As next-generation detectors extend the reach of observational capabilities, they also demand substantial theoretical progress in our understanding of classical two-body dynamics~\cite{Punturo:2010zz,LISA:2017pwj,Reitze:2019iox,Borhanian:2022czq,Purrer:2019jcp}.

A particularly promising approach has been the reformulation of the classical two-body problem within the framework of quantum field theory. In this setting, the \textit{post-Minkowskian} (PM) expansion -- where the dynamics are treated perturbatively in Newton's constant G -- has proven especially fruitful, facilitating the use of advanced techniques from collider physics to tackle gravitational interactions. In fact, remarkable progress has been achieved in the spinless sector, both through on-shell scattering amplitude methods~\cite{Cheung:2018wkq, Kosower:2018adc, Bern:2019nnu, Bern:2019crd, Cristofoli:2019neg, Bjerrum-Bohr:2019kec, Brandhuber:2021eyq, Bern:2021dqo, Bern:2021yeh, Damgaard:2023ttc} and worldline methods~\cite{Kalin:2020mvi, Kalin:2020fhe, kalin:2022hph, Dlapa:2023hsl, Dlapa:2021npj, Mogull:2020sak, Jakobsen:2021smu, Jakobsen:2022psy, Jakobsen:2023oow, Driesse:2024xad, Driesse:2024feo}. 
Extending these approaches to include the effects of spin, however, is essential for modelling real astrophysical systems. This challenge has spurred a surge of activity in both the amplitude-based and worldline-based approaches; see, e.g., Refs.~\cite{Bini:2017xzy, Bini:2018ywr,
  Vines:2017hyw, Vines:2018gqi, Guevara:2017csg, Guevara:2018wpp,
  Chung:2018kqs, Arkani-Hamed:2019ymq, Guevara:2019fsj, Chung:2019duq,
  Damgaard:2019lfh, Aoude:2020onz, Chung:2020rrz, Guevara:2020xjx,
  Bern:2020buy, Kosmopoulos:2021zoq, Chen:2021kxt,
  FebresCordero:2022jts, Bern:2022kto, Bern:2023ity, Menezes:2022tcs,
  Riva:2022fru, Damgaard:2022jem, Aoude:2022thd, Aoude:2022trd,
  Bautista:2022wjf, Gonzo:2023goe, Aoude:2023vdk, Lindwasser:2023zwo,
  Brandhuber:2023hhl, DeAngelis:2023lvf, Aoude:2023dui,
  Bohnenblust:2023qmy, Gatica:2024mur, Cristofoli:2021jas,
  Luna:2023uwd, Gatica:2023iws, Liu:2021zxr, Jakobsen:2021lvp,
  Jakobsen:2021zvh, Jakobsen:2022fcj, Jakobsen:2022zsx,
  Jakobsen:2023ndj, Jakobsen:2023hig, Heissenberg:2023uvo,
  Lindwasser:2023dcv, Bautista:2023sdf, Cangemi:2023ysz,
  Brandhuber:2024bnz, Chen:2024mmm, Correia:2024jgr, Bhattacharyya:2024kxj,
  Alaverdian:2024spu, Brandhuber:2024qdn, Brandhuber:2024lgl, 
  Akpinar:2024meg, Bohnenblust:2024hkw, Haddad:2024ebn,
  Bonocore:2024uxk, Akpinar:2025huz, Bohnenblust:2025gir}. 
  
The study of spinning binaries has a long and rich history within traditional perturbative approaches to general relativity, such as the Post-Newtonian (PN) expansion in the slow-velocity regime and the gravitational self-force (GSF) framework in the small mass-ratio limit. Powerful analytical and numerical techniques have been developed in these complementary perturbative regimes, often aided by the use of symmetries and the integrability of the probe motion in Kerr spacetime~\cite{Carter:1968ks,Carter:1968rr}. For the latter, the conservation of energy, azimuthal angular momentum and the Carter constant are usually introduced as a consequence of the presence of Killing vector~\cite{Dixon:1970zza} and tensor~\cite{Walker:1970un} symmetries of the background spacetime. Interestingly, such conservation laws have been extended to the case of a spinning probe in Kerr, through the introduction of the R\"udiger constant at linear in spin order~\cite{rudiger1981conserved,rudiger1983conserved} and a suitable extension of the Carter constant at the quadrupolar order~\cite{Compere:2023alp}; see Ref.~\cite{Compere:2021kjz} for a nice review.  It should be stressed that these conservation laws hold only for probes whose spin-induced multipole moments match those of a Kerr black hole, while for generic compact objects with generic Wilson coefficients -- most notably neutron stars -- integrability is expected to be broken already at quadratic order in spin.

Inspired by the simplicity of the Kerr black hole dynamics, and given the pivotal role of such symmetries in both the reformulation of the self-force program in action-angle variables~\cite{Schmidt:2002qk,Hinderer:2008dm} and the development of the effective-one-body (EOB) formalism for spinning binaries~\cite{Damour:1999cr,Damour:2001tu}, one might ask whether we can study this problem from the on-shell scattering amplitude perspective. This question is further motivated by the recent discovery of a hidden spin-shift symmetry -- wherein the amplitude remains invariant under a shift of the spin vector along the momentum transfer -- which was first identified at one-loop order~\cite{Bern:2022kto,Aoude:2022trd}, and later confirmed at the two-loop order~\cite{Akpinar:2024meg,Akpinar:2025bkt}. The fundamental origin of this symmetry, however, remains to be understood.

A natural setup to answer these questions is the Dirac bracket framework~\cite{Gonzo:2024zxo} (see also Refs.~\cite{Kim:2024grz, Kim:2024svw, Kim:2025hpn,Alessio:2025flu} for related studies), which allows one to efficiently compute the variation of dynamical variables in the two-body spinning phase space for scattering orbits -- including observables such as the impulse and spin kick -- directly from the gauge-invariant radial action extracted through the amplitude-action relation~\cite{Bern:2021dqo,Damgaard:2021ipf,Gonzo:2024zxo,Akpinar:2024meg,Akpinar:2025bkt}. In light of this, we will start by defining the conservation laws from the asymptotic scattering perspective, clarifying the relation with the usual conserved quantities in a Kerr spacetime. After extracting from the literature the relevant spinning radial actions in the probe limit and beyond -- up to 3PM at quartic in spin order -- we will then establish the validity of such symmetries, further comparing them with a new reformulation of the spin-shift symmetry in our phase space. Finally, we define a novel notion of asymptotic integrability, which we found to be satisfied by a spinning probe in Kerr to all PM orders and for generic spinning binaries at low PM orders.

\textit{Conventions---} 
In this \textit{Letter}, we work in the negative metric signature $(+-\ldots-)$ together with $\epsilon^{0123}=+1$ and $c = 1$. To reduce clutter, we adopt the shorthand notation 
$\hat{\delta}^{(d)}(x) = (2\pi)^d \delta^{(d)}(x)$ and $\hat{\mathrm{d}}^{d}k = \mathrm{d}^d k/(2\pi)^d$.

\section{Conserved quantities from the Dirac bracket formalism}\label{sec:conserved_DB}

In this section, we review the covariant Dirac bracket formalism discussed in Ref.~\cite{Gonzo:2024zxo} (see also Refs.~\cite{Kim:2024svw,Kim:2025hpn}) which we use to define conserved charges for the scattering of Kerr black holes. We begin by specifying the relevant kinematic phase space for the classical gravitational scattering of two spinning bodies, given by $\mathcal{H}_{\rm 2B} = \{v_1^{\mu},v_2^{\mu},s_1^{\mu}, s_2^{\mu}, b^{\mu}\}$. Here, $2\mathrm{B}$ stands for two-body, $v_i^\mu$ denotes the particle velocities, $s_i^\mu$ their normalized spin vectors, and $b^\mu = b_2^\mu - b_1^\mu$ is the initial impact parameter of the system. 
Then we consider the variation of an operator $\mathcal{Q}(v_i,s_i,b)$ within the KMOC formalism~\cite{Kosower:2018adc}
\begin{align}
 \Delta \hat{\mathcal{Q}} &= \hat{S}^{\dagger} \hat{\mathcal{Q}} \hat{S} - \hat{\mathcal{Q}} \nonumber \\
 &= \sum_{n = 1}^{\infty} \frac{(-i)^n}{\hbar^n n!} \underbrace{[\hat{N},[\hat{N}, \ldots,[\hat{N}}_{n \text{ times}}, \hat{\mathcal{Q}}]]]\,,
 \label{eq:operator_var}
\end{align}
where we have used the exponential representation of the S-matrix~\cite{Damgaard:2021ipf,Damgaard:2023ttc}
\begin{equation}
    \hat{S} = \exp\left(i \hat{N}/\hbar\right)\,.
\end{equation}
Focusing on the elastic $2 \to 2$ scattering of two massive particles, we parametrize the incoming and outgoing momenta as $p_1^\mu$, $p_2^\mu$ and $p_1^{\prime \mu}$, $p_2^{\prime \mu}$. In the classical limit, these are expanded as
\begin{align}
    &p_1^\mu=\bar{p}_1^\mu+\hbar \frac{\bar{q}^\mu}{2}, \quad p_1^{\prime \mu}=\bar{p}_1^\mu-\hbar \frac{\bar{q}^\mu}{2}, \nonumber \\
    &p_2^\mu=\bar{p}_2^\mu-\hbar \frac{\bar{q}^\mu}{2}, \quad p_2^{\prime \mu}=\bar{p}_2^\mu+\hbar \frac{\bar{q}^\mu}{2} \,,
\end{align}
where $\bar{p}_j^\mu = m_j v_j^\mu$ are classical momenta while $q = \hbar\bar{q}^\mu$ is the momentum transfer. The radial action is then defined via the Fourier transform of the elastic four-point matrix element of the operator $\hat{N}$ in the classical limit
\begin{align}
 &\hspace{-15pt}I_r(\{v_i,s_i,b\}) =  \int  \hat{\mathrm{d}}^{D} q \,\hat{\delta}\left(2 \bar{p}_1 \cdot q\right) \hat{\delta}\left(2 \bar{p}_2 \cdot q\right) \nonumber \\
 &\qquad \qquad \qquad \times \mathrm{e}^{i\left(q \cdot b\right) / \hbar} \langle p_1' p_2' |\hat{N}|p_1 p_2\rangle \Big|_{\hbar \to 0}\,,
 \label{eq:radial_act_def}
\end{align}
as proven rigorously in the spinless case~\cite{Damgaard:2023ttc}. Introducing the rapidity variable $y=v_1 \cdot v_2 > 1$ and retracing the connection between the radial action and the amplitude defined from the expansion $\hat{S} = 1 + i \hat{T}$, we arrive at the amplitude-action relation  
\begin{align}
    &i \mathcal{M}_4(q) = 4 \frac{m_1 m_2 \sqrt{y^2 -1}}{\hbar^2} \nonumber \\
    &\qquad \quad \times \int \mathrm{d}^{D-2} b \, e^{-i q \cdot b/\hbar}\, \left(e^{i I_r(b)/ \hbar} - 1 \right)\,,
    \label{eq:amplitude-radial}
\end{align}
which provides a prescription to obtain the radial action $I_r(b)$ from the four-point scattering amplitude $\mathcal{M}_4(q)$ for two massive spinning point particles by dropping certain divergent integrals after soft expansion\footnote{There are other equivalent ways to extract the radial action, either by integrating the scattering angle or the impulse~\cite{Damour:1999cr,Kalin:2019inp}, or by using conservation laws~\cite{Gonzo:2024zxo}.}~\cite{Bern:2021dqo, Bern:2021xze,Akpinar:2024meg,Akpinar:2025bkt}. 

The Dirac bracket formalism~\cite{Gonzo:2024zxo} provides the classical limit of commutators in the phase space $\mathcal{H}_{\rm 2B}$
\begin{align}
    [f(\cdot),g(\cdot)] \stackrel{\hbar \to 0}{\to} i \hbar \{f(\cdot),g(\cdot)\}_{\DB}\,,
\end{align}
so that the total variation of the expectation value $\langle \Delta \hat{\mathcal{Q}} \rangle$ \eqref{eq:operator_var} in the scattering process becomes, after inserting the two-body phase space completeness relation and Fourier transforming to impact parameter space,
\begin{align}
 \langle \Delta \hat{\mathcal{Q}} \rangle \Big|_{\mathrm{cons}} &= \sum_{n = 1}^{\infty} \frac{1}{ n!} \underbrace{\{I_r,\{I_r, \ldots,\{I_r}_{n \text{ times}}, \mathcal{Q}\}_{\DB}\}_{\DB}\}_{\DB}\,.
 \label{eq:operator_var_final}
\end{align}
This expression admits a natural extension to the dissipative sector upon including appropriate radiative kernels; see Ref.~\cite{Alessio:2025flu} for more details. However, in this work we focus exclusively on the conservative regime.
For the classical two-body dynamics of spinning particles with spin tensors $S_i^{\mu \nu}=m_i \epsilon^{\mu \nu}{ }_{\rho \sigma} v_i^\rho s_i^\sigma$, the Dirac brackets can be written as~\cite{Alessio:2025flu}
\allowdisplaybreaks{
\begin{subequations}
\label{eq:Dirac_brackets_final}
\begin{gather}
    \{b^\mu, v_j^\nu\}_{\DB} = - \frac{\mathrm{sgn}_j}{m_j} \Pi^{\mu\nu}, \\
    \{b^\mu, s_j^\nu\}_{\DB} = \frac{\mathrm{sgn}_j}{m_j}
\Pi^{\mu\rho}s_{j\rho}v_j^\nu, \\
    \{b^\mu, b^\nu\}_{\DB} = \frac{b^\mu}{y^2-1}\left[ \frac{v_1^\nu-yv_2^\nu}{m_1} +\frac{yv_1^\nu-v_2^\nu}{m_2} \right]-(\mu\leftrightarrow\nu) \nonumber \\
    \quad+\Pi^{\mu\rho} \Pi^{\nu\sigma} \left[ \frac{S_{1\rho\sigma}}{m_1^2} +\frac{S_{2\rho\sigma}}{m_2^2} \right],\\
    \{s_i^\mu, s_j^\nu\}_{\DB} = \delta_{ij}  \frac{S_i^{\mu\nu}}{m_i^2}, 
\end{gather}
\end{subequations}
}
where we have defined the projectors and the sign factors
\begin{gather}
    \mathrm{sgn}_1 := -1, \quad 
    \mathrm{sgn}_2 := +1,  \\ 
    \mathcal{P}_1^{\mu\nu} := \frac{v_1^\mu (v_1^\nu - y v_2^\nu)}{y^2 -1},\quad 
    \mathcal{P}_2^{\mu\nu} := \frac{v_2^\mu (v_2^\nu - y v_1^\nu)}{y^2 -1}, \\
    \Pi^{\mu\nu}:=\eta^{\mu\nu}+\mathcal{P}_1^{\mu\nu}+\mathcal{P}_2^{\mu\nu}.
    \label{eq: transverse-projector}
\end{gather}
As such, given a quantity $\mathcal{Q}(\{v_i, s_i, b\})$ in our phase space $\mathcal{H}_{\rm 2B}$, we say that such a quantity is conserved if 
\begin{align}\label{eq:vanishing_bracket}
    \{\mathcal{Q}, I_r\}_{\DB} = 0 \rightarrow  \langle \Delta \mathcal{Q} \rangle = 0\,,
\end{align}
just as we would do in a Hamiltonian formulation\footnote{Unlike the usual Hamiltonian flow, the Dirac brackets used here are gauge-invariant. It should be possible to extend this definition of conserved quantity to the generic dissipative case upon inclusion of the radiative kernels, as suggested in Ref.~\cite{Alessio:2025flu}.}.

\subsection*{Asymptotic vs local conservation laws}

Now we review how exact local conservation laws~\cite{Compere:2021kjz,Compere:2023alp} can be formulated in the probe limit for a spinning test particle in a Kerr background at quadrupolar order, and then discuss their connection to the vanishing of the Dirac brackets in the asymptotic on-shell expansion \eqref{eq:vanishing_bracket}. 

The Kerr spacetime admits two Killing vectors, $\xi_t^\mu = (\partial_t)^\mu$ and $\xi_\phi^\mu = (\partial_\phi)^\mu$, associated respectively with stationarity and axisymmetry, as well as a rank-2 Killing tensor $K_{\mu \nu} = Y_{\mu\rho} Y_{\nu}{}^{\rho}$, constructed from an antisymmetric Killing-Yano tensor $Y_{\mu\nu} = -Y_{\nu\mu}$ that encodes a hidden symmetry of the background. These geometric structures give rise to all the conserved quantities for the motion of a spinning test particle in the quadrupolar approximation, and -- as we will show -- also beyond~\cite{rudiger1981conserved,rudiger1983conserved,Compere:2021kjz}.

We define the covariant four-momentum for the probe particle as $\pi_\mu = p_\mu -  \omega_{\mu ab} S^{ab}/2$, where $S^{ab}$ is the spin tensor in the local Lorentz frame and $\omega_{\mu ab}$ is the spin connection, ensuring local Lorentz invariance in curved spacetime. The symmetries of Kerr then yield four conserved quantities. First, the Killing vectors lead to spin-dependent energy and angular momentum~\cite{rudiger1981conserved,rudiger1983conserved}:
\begin{align}
E &= -\pi_\mu \xi_t^\mu + \frac{1}{2} S^{\alpha \beta} \nabla_\beta \xi_{t\alpha}, \\
L &= \pi_\mu \xi_\phi^\mu - \frac{1}{2} S^{\alpha \beta} \nabla_\beta \xi_{\phi\alpha}.
\end{align}
Second, the Killing tensor gives rise to the spin-corrected Carter constant~\cite{Carter:1968ks,Compere:2021kjz,Compere:2023alp}: 
\begin{align}
\label{eq:carterQ}
Q &= Y_{\mu \rho} Y^\rho{}_\nu p^\mu p^\nu + 4 \xi_t^\lambda \epsilon_{\lambda \mu \sigma[\rho} Y_{\nu]}{}^\sigma S^{\mu \nu} p^\rho \\
&\quad - \Big[ g_{\mu \rho} \left( (\xi_t)_\nu (\xi_t)_\sigma - \frac{1}{2} g_{\nu \sigma} \xi_t^2 \right) \nonumber \\
& \qquad - \frac{1}{2} Y_\mu{}^\lambda \left( Y_\rho{}^\kappa R_{\lambda \nu \kappa \sigma} + \frac{1}{2} Y_\lambda{}^\kappa R_{\kappa \nu \rho \sigma} \right) \Big] S^{\mu \nu} S^{\rho \sigma}\,. \nonumber
\end{align}
Finally, there is another hidden symmetry which yields a conserved projection of the spin along the principal plane of the Kerr geometry, the R\"udiger linear invariant:
\begin{align}
Q_Y =\frac{1}{2} \epsilon^{\mu \nu \alpha \beta} S_{\mu \nu} Y_{\alpha \beta} \,.
\end{align}
These four quantities -- $E$, $L$, $Q$ and $Q_Y$ -- fully characterize the integrable structure of spinning particle dynamics in Kerr at quadratic order in spin~\cite{Compere:2023alp}. 

In order to make a connection with the asymptotic radial action, we evaluate these four quantities in the limit $r \to +\infty$.
In this limit, denoting for simplicity their asymptotic forms with the same symbols, the conserved quantities reduce to expressions that depend only on the two-particle kinematics~\cite{Gonzo:2024zxo,Vinesslides}:
%
\begin{align}
    \label{eq:conserved-quantities-asymptotic}
    \hspace*{-0.2cm} \frac{E}{m_1}&=y,\\
    \hspace*{-0.2cm} \tilde{L}&=l\cdot s_2 -y (s_1\cdot s_2)+(v_2 \cdot s_1)(v_1 \cdot s_2)\,,\label{eq:asymptotic-L}\\[0.1cm]
    \hspace*{-0.2cm} Q_Y&= l\cdot s_1 + y (s_1\cdot s_2) - (v_2 \cdot s_1)( v_1 \cdot s_2)\,,\\[0.2cm]
    \hspace*{-0.2cm} Q&=  -l^2 -2 y (l\cdot s_2) - (y^2-1)s_2^2 \\
    & - (v_1\cdot s_2)^2 +2y (l\cdot s_1) +2 (y^2+1)(s_1\cdot s_2)\nn\\
    &-2y (v_1\cdot s_2)(v_2 \cdot s_1) - (y^2-1)s_1^2-(v_2\cdot s_1)^2\,, \nn
\end{align}
where we define $l^\mu = \epsilon^{\mu \nu \rho \sigma} b_{\nu}v_{1\rho}v_{2\sigma}$, $m_1 \tilde{L}=|s_2| L$ and choose particle 1 as the probe while treating particle 2 as the background. Here $E$ and $L$ are the energy and axial angular momentum of particle $1$ in the rest frame of $2$, respectively, while $Q_Y$ is the linear R\"udiger invariant and $Q$ is the quadrupolar extension of the Carter constant.  

A word of caution is necessary here regarding the distinction between local and asymptotic conservation laws in the sense of \eqref{eq:vanishing_bracket}: while the former implies the latter, the converse need not hold. In particular, starting at quadratic order in spin, curvature-dependent contributions generate subleading corrections in the $1/r$ expansion to the asymptotic quantities in \eqref{eq:conserved-quantities-asymptotic}. For example, the kinematic mass $m_1^2 = \eta_{\mu \nu}p_1^\mu p_1^\nu$ is not conserved off-shell at $\mathcal{O}(S_1^2)$ in the worldline setup. The conserved quantity at this order is instead the dynamical mass~\cite{Jakobsen:2021zvh}
\begin{equation}
    \mu_1^2 = m_1^2 - \frac{1}{4} R_{\mu\nu\rho\sigma}\, S_1^{\mu\nu} S_1^{\rho\sigma}
    + \mathcal{O}(S_1^3),
\end{equation}
whose curvature-dependent correction vanishes in the $r\to\infty$ limit, effectively giving $\mu_1 \to m_1$ in our on-shell approach. Similar curvature terms appear on the last line of the quadratic-in-spin correction to the Carter constant~\eqref{eq:carterQ}, but likewise drop out asymptotically.\footnote{These terms become important when extending the conservation laws into the bulk or describing the local off-shell dynamics in the strong field regime, where curvature effects cannot be discarded.} Nevertheless, since our primary goal is to analyse the simplicity of the radial action and other on-shell quantities relevant for Kerr black hole scattering, the asymptotic formulation adopted here provides precisely the gauge-invariant notion of conservation laws appropriate for our purposes.
%


\begin{table*}[!ht]
    \centering
    \begin{minipage}{0.35\textwidth}
    \centering
    \begin{tabular}{@{}|P{1.5cm}|c|c|c|c|c|@{}}
    	\hline \multicolumn{6}{|c|}{\textbf{Probe limit}} \\ \hline
        \toprule
        $\tilde{L}, Q_Y, Q$ & $\cO(s^2)$ & $\cO(s^3)$ & $\cO(s^4)$ & $\cO(s^5)$ & $\cO(s^\infty)$\\ \hline
        \midrule
        $G^1$ & \tickg & \tickg & \tickg & \tickg &\tickg\\ \hline
        $G^2$ & \tickg & \tickg & \tickg & \crossr & ?\\ \hline
        $G^3$ & \tickg & \tickg & \tickg & ? &? \\ \hline
        $G^4$ & \tickg & \tickg & ? & ? &?\\ \hline
        $G^5$ & \tickg & ? & ? & ? &? \\ \hline
        \bottomrule
    \end{tabular}
    \end{minipage}
    \hspace{0.4cm}
    \begin{minipage}{0.55\textwidth}
    \centering
    \begin{tabular}{@{}|P{1.3cm}|c|c|c|c|c|c|c|@{}}
    	\hline \multicolumn{8}{|c|}{\textbf{Beyond probe limit (conservative)} } \\ \hline
        \toprule
        $\tilde{L}, Q_Y, Q$ & $\cO(s^1)$ & $\cO(s_1^2 s_2^0)$ &$\cO(s_1^3 s_2^0)$ & $\cO(s_1^4 s_2^0)$ & $\cO(s_1^5 s_2^0)$ & $\cO(s_1^\infty s_2^0)$ & $\cO(s_1^1 s_2^1)$ \\ \hline
        \midrule
        $G^1$ & \tickg & \tickg & \tickg & \tickg &\tickg & \tickg & \crossr \\ \hline
        $G^2$ & \tickg &\tickg & \tickg & \tickg & \crossr & ? & \crossr \\ \hline
        $G^3$ & \tickg & \crossr &\crossr & \crossr & ? & ? & \crossr \\\hline
        \bottomrule
    \end{tabular}
    \end{minipage}
    \caption{Conservation results in the probe limit (left) and beyond the probe limit (right), restricted to the conservative sector. A `?' indicates orders in the radial action that are currently unknown. In the probe limit, these conservation laws enhance to asymptotic integrability. The pattern shown applies to all quantities $\tilde{L}, Q, Q_Y$ (and similarly to the $s_1 \leftrightarrow s_2$ contributions beyond the probe limit), with the sole exception of the $\mathcal{O}(G^2 s^5)$ term in the probe limit which is discussed further in the Supplemental Material. We find that spin-shift symmetry holds precisely in the same regimes in which integrability is realised, both in the probe limit and beyond.}
    \label{tab:core_conservation_results}
\end{table*}

\section{Unexpected symmetries of Kerr black hole scattering}
\label{sec:symmetries}
In this section, we test the symmetries of the radial action using data available in the literature:
\begin{itemize}
    \item 1PM and all orders in spin~\cite{Guevara:2019fsj}, 
    \item 2PM through $\cO(s^{11})$~\cite{Bohnenblust:2024hkw}\footnote{It is still unknown if the dynamics at $\mathcal{O}(s^{5})$ and beyond reproduces the dynamics of a Kerr black hole.}, 
    \item 2PM and 3PM through $\cO(s^2)$~\cite{Jakobsen:2022fcj}, 
    \item 2PM and 3PM through $\cO(s_1^4 s_2^0)$~\cite{Akpinar:2025bkt}, 
    \item $\infty$PM through $\cO(s_1^1 s_2^\infty)$ in the probe limit~\cite{Gonzo:2024zxo}\footnote{The results of Ref.~\cite{Gonzo:2024zxo} are already written in terms of conserved quantities and so automatically obey \eqref{eq:vanishing_bracket}.},
    \item 3PM through $\cO(s^4)$, 4PM through $\cO(s^3)$ and 5PM through $\cO(s^2)$ in the probe limit~\cite{Hoogeveen:2025tew}, 
\end{itemize}
where the total power in spin is denoted by $\cO(s^n)$ (for example, $\cO(s^2)=\cO(s_1^2)+\cO(s_1 s_2)+\cO(s_2^2)$). For Refs.~\cite{Jakobsen:2022fcj,Hoogeveen:2025tew}, we extracted the radial action uniquely via an ansatz matched to their impulse. We have checked that the radial actions listed above are in agreement where they overlap (up to metric signature and Levi-Civita conventions), and we have included their expression in the ancillary file of this \textit{Letter} for completeness.

Recall from \eqref{eq:vanishing_bracket} that the radial action exhibits an (asymptotic) symmetry if its Dirac bracket with the corresponding conserved quantity vanishes. Since available results hold only up to a fixed order in the spins, conservation is likewise expected only at that order. More specifically, for a conserved quantity $\cQ$ from the previous section and a radial action $I_r^{n,m}$ including terms up to $\cO(s_1^n s_2^m)$, conservation requires
\begin{equation}\label{eq:vanishing_bracket_spin}
    \{\mathcal{Q}, I_r^{n,m}\}_{\DB} = 0 +\cO(s_1^{n+1}s_2^{m})+\cO(s_1^{n}s_2^{m+1})\,.
\end{equation}
The Dirac brackets do not mix loop orders but do mix spin orders, so all spin terms up to the desired order must be included in $I_r^{n,m}$. Moreover, because the brackets involve explicit factors of $1/m_i$, they also mix powers in the self-force expansion. In fact, in the probe limit $m_2\rightarrow \infty$, one must therefore use the probe limit of the radial action and discard any terms in \eqref{eq:Dirac_brackets_final} proportional to $1/m_2$.

Our main results are summarized in Table \ref{tab:core_conservation_results}, but there are some general features that we wish to mention. First, in the probe limit 
$m_2\rightarrow\infty$: 
\begin{itemize}
    \item The $\cO(G^{\infty} s_1^1  s_2^\infty)$ radial action commutes with all conserved quantities.
    \item The 1PM radial action commutes with all of the conserved quantities to all orders in spin.
    \item The 2PM and 3PM radial action commutes with all conserved quantities through $\cO(s^4)$; likewise for the 4PM radial action through $\mathcal{O}(s^3)$ and the 5PM radial action through $\mathcal{O}(s^2)$.
\end{itemize}
%
Beyond the probe limit, the conservation laws continue to hold for the spin-orbit sector -- that is, for terms of order $\mathcal{O}(s_{1}^{0}s_{2}^{1})$ or $\mathcal{O}(s_{1}^{1}s_{2}^{0})$ -- since these couplings are universal. However, once we go beyond the spin-orbit level and include, for example, $\mathcal{O}(s_1^1 s_2^1)$ contributions, the conservation laws are generically broken.
We also discuss further checks of conservation laws for terms beyond $\cO(s_1^n s_2^m)$ for $n+m \leq 4$ and including non-minimal couplings in the supplemental material.

\subsection*{Reinterpreting spin-shift symmetry}

It is well-known that the spinning four-point amplitude exhibits a so-called \textit{spin-shift symmetry} through quartic order in spin up to two-loop order~\cite{Bern:2022kto,Aoude:2022trd,Akpinar:2024meg,Akpinar:2025bkt}. This means that the amplitude is invariant under the shift
\begin{equation}
    s_i^\mu \rightarrow s_i^\mu + \xi q^\mu\,,
    \label{eq: spin-shift-symmetry}
\end{equation}
for any $\xi$ with a formal counting of $\mathcal O(1/|q|^2)$. This statement may be realized at the linearized level through the action of a differential operator
\begin{equation}
    \hat{D}_{\mathrm{SS}} = q^\mu \frac{\partial}{\partial s_i^\mu}\,.
\end{equation}
In position space, the statement \eqref{eq: spin-shift-symmetry} becomes
\begin{equation}
    \hat{O}_{\mathrm{SS}} I_r(\{v_i, s_i, b\}) = 0\,, \quad \hat{O}_{\mathrm{SS}} = \Pi_{\mu \nu}\frac{\partial}{\partial b^\mu}\frac{\partial}{\partial s_i^\nu}\,,
    \label{eq:spin-shift_bspace}
\end{equation}
where we use the transverse projector in \eqref{eq: transverse-projector},
%
%
since the Fourier transform is performed in the transverse subspace. Importantly, we have checked that the radial action satisfies \eqref{eq:spin-shift_bspace} when the spin-shift symmetry \eqref{eq: transverse-projector} is exhibited by the corresponding spinning four-point amplitude in momentum space. We have also verified that the probe limit radial action in Ref.~\cite{Gonzo:2024zxo} at $\cO(G^{\infty} s_1^1  s_2^\infty)$ is invariant under such a symmetry, as the radial momentum $p_r$ itself is invariant.

Now we must understand if $\hat{O}_{\mathrm{SS}}$ can be generated by any scalar function on the phase space. In classical mechanics, physical symmetries correspond to Hamiltonian vector flows generated by scalar charges, say $\chi$, acting via the Poisson or Dirac bracket, i.e. $\mathcal{V}_\chi(f) = \{f,\chi\}$. These vector fields necessarily involve at most one derivative with respect to phase space coordinates, which are realized by the action of the bracket. However, any transformation involving multiple derivatives cannot live in the space of Hamiltonian vector fields. Thus, $\hat{O}_{\mathrm{SS}}$ cannot be interpreted as the flow generated by a conserved charge. Instead this reflects a field-space redundancy of the theory, analogous to gauge invariances, which leave physical observables unchanged but do not correspond to physical symmetries associated to Noether currents. In fact, to draw a useful analogy, if we consider scattering amplitudes involving massless vector or graviton states of momentum $k$ and polarization $\epsilon(k)$, then physical observables are invariant under a shift in the polarization
\begin{equation}
    \epsilon^\mu(k) \rightarrow \epsilon^\mu(k) + \zeta k^\mu\,,
\end{equation}
for any $\zeta$. This reflects a gauge redundancy in the polarization vector, rather than a symmetry of the Hilbert space, and thus no Noether charge is associated to this.

We leave a more rigorous analysis of the nature of the spin-shift symmetry to future work. 

\section{The integrable sector of Kerr black hole dynamics}\label{sec:integrable_dyn}

Here we define a new asymptotic notion of classical integrability for the scattering dynamics of binary black holes. The dynamics of the spinning radial action $I_r$ is determined by five vectors $\{v_1^{\mu},v_2^{\mu},s_1^{\mu},s_2^{\mu},b^{\mu}\}$, corresponding to a $5\times 4=20$ dimensional phase space after using translational freedom to choose $b^{\mu} = b_2^{\mu} - b_1^{\mu}$. We then impose on-shell conditions $v_i^2 = 1$, the covariant spin supplementary condition (SSC) $v_i \cdot s_i = 0$, and transversality $b \cdot v_i = 0$. These conditions provide $6$ constraints, thereby reducing the phase space to $14$ dimensions. This is the space in which the Dirac brackets \eqref{eq:Dirac_brackets_final} are defined, but we may define additional conserved quantities to further constrain the phase space. 

First, we consider the conservation of spin magnitudes $\left\langle \Delta s_i^2 \right\rangle\big|_{\mathrm{cons}} = 0$, which follows directly from the bracket structure since $s_i^2$ are Casimirs of the algebra. This yields two additional constraints, further reducing the phase space to $14 - 2 = 12$ dimensions. Translation invariance implies the conservation of center of mass (COM) momentum
\begin{align}
    v_{\rm CM}^\mu := \frac{m_1 v_1^\mu + m_2 v_2^\mu}{M}\,, \qquad M := m_1 + m_2\,,
\end{align}
so that $\left\langle \Delta v_{\rm CM}^\mu \right\rangle\big|_{\mathrm{cons}} = 0$. It is convenient to introduce the \emph{relative momentum}
\begin{align}
    p_{\rm rel}^\mu := \frac{E_2 p_1^\mu - E_1 p_2^\mu}{M}\,,
\end{align}
which reduces to $(0, \vec{p}_\infty)$ in the COM frame. Since $y$, $E_1$, $E_2$, and $|\vec{p}_\infty| \propto \sqrt{y^2 - 1}$ are also conserved under Dirac bracket evolution, the magnitude of $\vec{p}_{\rm rel}$ is fixed. Writing $(\vec{v}_1,\vec{v}_2)$ in terms of $(\vec{v}_{\rm CM},|\vec{p}_{\infty}| \vec{u}_{\rm rel})$, where $\vec{u}_{\rm rel} = \vec{p}_{\rm rel} / |\vec{p}_{\infty}|$ is a unit vector, implies four additional reductions, yielding a final phase space of dimension $12-4 = 8$.

Therefore we obtain the reduced phase space $\mathcal{H}^{\rm red}_{\rm 2B}$ 
\begin{align}
    \label{eq:phase-space_fin}
    &\hspace{0.8cm}\mathcal{H}^{\rm red}_{\rm 2B} = \left\{ \vec{b}_{\perp}, \vec{u}_{\rm rel}, \vec{s}_{1\perp}, \vec{s}_{2\perp} \right\}\,, \\
    &b^{\mu}_{\perp} = \Pi^{\mu \nu}b_{\nu}\,, \quad s_{i\perp}^\mu = \left( \eta^{\mu\nu} - v_i^\mu v_i^\nu \right) \frac{s_{i\nu}}{s_i^2}\,.\nonumber
\end{align}
This dynamical system, with a $2n$-dimensional reduced phase space equipped with a Dirac bracket structure, is said to be asymptotically integrable in the Liouville sense\footnote{In our case, although there is no Hamiltonian in the usual sense, the radial action generates a flow on the reduced phase space that plays an analogous role to a Hamiltonian vector field. It would be interesting to place this concept on a rigorous footing.} if there exist $n$ independent phase-space functions $F_i$ -- said to be in \textit{involution} -- that are all mutually commuting under the Dirac brackets, i.e. $\{F_i, F_j\}_{\DB} = 0$ for all $i,j$. This notion is \emph{generically weaker} than bulk Liouville integrability: the existence of commuting asymptotic charges does not guarantee that the full equations of motion admit the same number of conserved quantities in involution. As seen in the spinning-probe dynamics in Kerr~\cite{Witzany:2019nml,Gonzo:2024zxo,Witzany:2024ttz,Skoupy:2024uan,Ramond:2022vhj,Ramond:2024sfp,Ramond:2024ozy} and discussed earlier, the obstruction arises from curvature-dependent spin couplings that are absent in the asymptotic on-shell regime but are present in the local bulk dynamics, breaking the correspondence between asymptotic and local conservation laws.


While the reduced phase space \eqref{eq:phase-space_fin} is the natural one for the conservative dynamics of spinning black hole binaries, we showed 
earlier that there exists further conservation laws, at least for a spinning probe particle in Kerr. Using their asymptotic expansion \eqref{eq:conserved-quantities-asymptotic}, we define 
\begin{align}
    \vec{F}^{\rm BH} = \{I_r,\tilde{L},Q,Q_Y\} \,,
\end{align}
which we use to test integrability of our spinning radial action. 
Within the probe limit $m_2\rightarrow\infty$ when both black holes are spinning, we have checked that
%
\begin{equation}
    \{F^{\rm BH}_i, F^{\rm BH}_j\}_{\DB} = 0 \quad \forall \quad  i,j=1,2,3,4\;.
\end{equation}
The above relation holds in the probe limit at the same orders in PM and spin where conservation holds in Table~\ref{tab:core_conservation_results}.
%
Beyond the probe limit, when both black holes are spinning, $\tilde{L}$, $Q$ and $Q_Y$ are not in involution and so integrability breaks down (albeit at least at 1PM this can be restored; see the Supplemental Material). In the case of one spinning black hole, say, $s_1\rightarrow0$ ($s_2\rightarrow0$) we find that $Q_Y$ ($\tilde{L}$) is zero, such that the remaining bracket $\{Q,\tilde{L}\}$ ($\{Q,Q_Y\}$) vanishes \textit{without taking the probe limit}. We therefore conclude: 

\begin{itemize}
    \item (Conjecture) The probe dynamics of one spinning black hole (BH) in the background of a Kerr BH is asymptotically integrable at all PM orders up to quartic order in spin for the probe;
    \item The scattering of one spinless BH and one spinning BH is asymptotically integrable at the 2PM order and quartic order in the BH spin, even beyond the probe limit.
\end{itemize}
This property can be made manifest by expressing the radial action in terms of the conserved charges $\tilde{L}$,$Q$ and $Q_Y$. These results are included in the ancillary file.

\section{Bootstrapping the spinning radial action from symmetries}\label{sec:bootstrap}
%

An intriguing consequence of our analysis is that integrability and spin-shift symmetry in the probe limit impose nontrivial constraints on the structure of the radial action, enabling us to bootstrap the radial action under the assumption that both properties hold.

Consider a generic expression for the spinning radial action at order $G^n$, including terms up to $s_1 + s_2 \leq 4$:
\begin{align}
    \mathcal{I}^{(n)}_{r}(b) = \frac{1}{|b|^{(n-1)(1-2\epsilon)}}\sum_{s_1,s_2}\sum_{i}\alpha^{(s_1,s_2,i)}\mathcal{J}^{(s_1,s_2,i)}\,,
\end{align}
where $|b|^2 = -b^2$ and $\epsilon = (4 - d)/2$ is the usual dimensional regularization parameter. The integers $s_1$ and $s_2$ label the spin orders of the two bodies, while $i$ indexes independent structures at each spin level. The coefficients $\alpha$ are general functions of the kinematic variables, and $\mathcal{J}$ represent the corresponding independent spin structures.

At this order in spin, there are a total of 70 independent spin structures, and thus 70 a priori undetermined coefficients in the radial action. The objective is to constrain these coefficients by imposing conservation laws in the probe limit and enforcing spin-shift symmetry:
\begin{itemize}
    \item Conservation of the charges $Q_Y$ and $Q$ yields 40 independent constraints;
    \item Conservation of the azimuthal angular momentum $\tilde{L}$ imposes no further constraints, as the transversality conditions $v_i \cdot b = 0$ already fix the preferred plane. Consequently, $\tilde{L}$ is automatically conserved for the radial action in the probe limit;
    \item Spin-shift symmetry imposes 15 constraints.
\end{itemize}
In total, this reduces the number of free parameters to 15. Remarkably, this exactly matches the independent coefficients appearing in the aligned-spin scattering angle. As a result, knowledge of the aligned-spin scattering angle suffices to determine all 70 coefficients in the general spinning radial action. Notice that this is reminiscent of the tutti-frutti method~\cite{Bini:2019nra,Bini:2020wpo}, albeit using symmetries in place of combining different perturbative schemes.

\section{Conclusions and future directions}\label{sec:conclusion}

The study of symmetries in the context of classical gravitational two-body interactions has been instrumental for developing new analytical techniques in traditional general relativity, from the self-force to the effective one-body theory. In this \textit{Letter}, we have developed a new efficient on-shell method based on the Dirac bracket formalism to study conservation laws in the conservative scattering scenario, using only the spinning gauge-invariant radial action extracted from various PM calculations.

First, we derived the asymptotic expansion of the exact conserved quantities in the probe limit -- energy $E$, azimuthal angular momentum $\tilde{L}$, R\"udiger invariant $Q_Y$ and quadrupolar Carter constant $Q$ -- for a spinning particle in Kerr spacetime, expressing them in terms of kinematic invariants. Second, we have proposed a new on-shell definition of asymptotic Liouville integrability inspired by the Hamiltonian case, where charges are in involution with the Dirac bracket framework.

Our results for the conservation laws are summarized in Table \ref{tab:core_conservation_results}. We find that, surprisingly, $\{E,\tilde{L},Q,Q_Y\}$ are conserved and in involution up to quartic order in spin for the spinning probe particle in Kerr, implying that our notion of integrability holds beyond the quadratic in spin results of Ref.~\cite{Compere:2023alp}. Moreover, the same happens for generic spinning-spinning scattering at 1PM order and spinless-spinning scattering at 2PM order. Interestingly, imposing asymptotic integrability allows one to drastically simplify the bootstrap of generic spinning interactions for the radial action, implying in particular that the dynamics is fully determined by the aligned-spin case.  

Along these lines, another type of symmetry -- known as spin-shift -- has been found for amplitudes in momentum space, whose origin has not been clarified, accompanying the symmetries associated with conservation laws. By translating this symmetry as an operator in position space, we notice that this cannot be interpreted as a standard spacetime symmetry, but more like a (redundant) gauge symmetry, which deserves a further investigation.

Summarizing, our work indicates that the Kerr dynamics is even more constrained than previously anticipated, revealing a richer underlying geometric structure. One pressing question is to understand what are the implications for the integrability of the quartic in spin probe in Kerr for the bound dynamics, perhaps by finding a suitable coordinate and tetrad basis where the equations of motion are separable~\cite{Skoupy:2024uan}. Another tantalizing prospect would be to extend our findings to the dissipative case, reformulating the evolution equations for $\{\dot{E},\dot{\tilde{L}},\dot{Q},\dot{Q}_Y\}$ used in self-force theory~\cite{Hinderer:2008dm} in terms of the phase space integrals involving higher-point kernels~\cite{Alessio:2025flu}. Finally, we look forward to further applications of our on-shell notion of conservation laws and asymptotic integrability beyond the gravitational case.

\textbf{\textit{Acknowledgments---}}
We thank Fabian Bautista, Geoffrey Comp\`ere, Ricardo Monteiro, Paul Ramond, Leo Stein and Vojtech Witzany for useful discussions. 
We thank the authors of~\cite{Bohnenblust:2024hkw} for sharing an early revision of their ancillary files.
D.A.\ is supported by an STFC studentship.
The work of G.R.B.\ is supported by the U.K. Royal Society through Grant URF\textbackslash R1\textbackslash 20109.
The work of R.G.\ is supported by the Royal Society grant RF\textbackslash ERE\textbackslash 231084.
M.Z.’s work is supported in part by the U.K.\ Royal Society through
Grant URF\textbackslash R1\textbackslash 20109.
For the purpose of open access, the authors
have applied a Creative Commons Attribution (CC BY) license to any
Author Accepted Manuscript version arising from this submission.

\bibliography{hiddenSymmetries.bib}
\newpage
\appendix
\onecolumngrid
\newpage
\begin{center}
\textbf{Supplemental Material}
\end{center}

\section{Additional conservation checks}
Here we list additional checks of asymptotic integrability using the Dirac brackets and the radial actions discussed in the main text, including terms beyond $\cO(s_1^n s_2^m)$ for $n+m \leq 4$ and non-minimal couplings. 

At 2PM, the results of Ref.~\cite{Bohnenblust:2024hkw} go beyond fourth order in the probe spin and contain results through $\cO(s^{11})$. For these we find the following pattern of conservation in the probe limit $m_2\rightarrow\infty$, where n/a denotes cases in which the relevant conserved quantity is zero due to one of the BHs being spinless.

 \begin{center}
	\begin{tabular}{|c|c|c|c|c|c|c|}
		\hline
        \parbox[c][0.9cm][c]{3.5cm}{\centering 
        Probe limit: 2PM~\cite{Bohnenblust:2024hkw} \\ \small (Minimal coupling)
        } 
        & $\cO(s^4)$ & $\cO(s_1^5 s_2^0)$ & $\cO(s_1^{11}s_2^0)$ & $\cO(s_1^4 s_2^7)$ & $\cO(s_1^5s_2^1)$ &$\cO(s^{11})$\\ \hline
		$\tilde{L}$ & \tickg & n/a & n/a & \tickg& \tickg&\tickg \\ \hline
		$Q_Y$ & \tickg & \crossr & \crossr & \tickg & \crossr&\crossr\\ \hline
		$Q$ & \tickg & \tickg & \tickg & \tickg & \crossr &\crossr \\ \hline
	\end{tabular}
\end{center}
These results hold regardless of whether terms involving $|s_i|$ are included in the radial action. Beyond the probe limit the conservation is broken already at the orders in spin shown in the main text. 

We have also extracted the radial action from Ref.~\cite{Jakobsen:2022fcj} including non-minimal spin coupling terms at $\cO(s^2)$ used to describe e.g. neutron stars. The results of our conservation checks are given below, where an asterisk indicates the inclusion of non-minimal couplings. 
\begin{center}
	\begin{tabular}{|c|c|c|c|c|c|}
		\hline
        \parbox[c][0.9cm][c]{3.5cm}{\centering 
        Probe limit: 3PM~\cite{Jakobsen:2022fcj} \\ \small (Non-minimal coupling)
        } 
        & $\cO(s_1^1s_2^0)$ & $\cO(s_1^0s_2^1)$ & $\cO(s_1^1s_2^1)$ & $\cO(s_1^2s_2^0)^\ast$ & $\cO(s_1^0s_2^2)^\ast$ \\ \hline
		$\tilde{L}$ & n/a & \tickg & \tickg & n/a & \tickg\\ \hline
		$Q_Y$& \tickg & n/a & \tickg & \crossr & n/a\\ \hline
		$Q$ & \tickg & \tickg & \tickg & \tickg& \crossr \\ \hline
	\end{tabular}
\end{center}
This table applies to the results at 1PM, 2PM or 3PM since they exhibit the same pattern of conservation. These findings seem to be consistent with the results of Ref.~\cite{Compere:2023alp} on the (non-)existence of conserved quantities for generic spinning test bodies in a fixed Kerr background. We find that demanding conservation of the charges enforces that the Wilson coefficients of the non-minimal couplings to be zero. Note that, as observed in the main text, the azimuthal angular momentum is always conserved in the probe limit. 

Furthermore, we have also considered the conservation of non-minimal terms beyond the probe limit. As before, the same pattern applies to 1PM, 2PM or 3PM.
\begin{center}
	\begin{tabular}{|c|c|c|c|c|c|}
		\hline
         \parbox[c][0.9cm][c]{4.8cm}{\centering 
        Beyond probe limit: 3PM~\cite{Jakobsen:2022fcj} \\ \small (Non-minimal coupling)
        } 
        & $\cO(s_1^1s_2^0)$ & $\cO(s_1^0s_2^1)$ & $\cO(s_1^1s_2^1)$ & $\cO(s_1^2s_2^0)^\ast$ & $\cO(s_1^0s_2^2)^\ast$ \\ \hline
		$\tilde{L}$ & n/a & \tickg & \crossr & n/a & \crossr\\ \hline
		$Q_Y$& \tickg & n/a & \crossr & \crossr & n/a\\ \hline
		$Q$ & \tickg & \tickg & \crossr & \crossr& \crossr \\ \hline
	\end{tabular}
\end{center}

\section{The Carter constant from the Newman-Janis shift}
%
Here we show how one may obtain the Carter constant through a Newman-Janis shift. Our conventions are the same as in the main text. Consider the scattering of a spinless probe in the background of a Schwarzschild black hole. The conserved angular momentum, defined for asymptotic incoming and outgoing states, is
\begin{equation}
    l^\mu = \epsilon^{\mu \nu \rho \sigma}b_{\nu}v_{1\rho}v_{2\sigma}\,,
\end{equation}
with conserved squared norm $l^2 = (y^2-1)b^2$. It is a well-established result that the Newman–Janis shift transforms the Schwarzschild metric into the Kerr metric~\cite{Newman:1965,Drake:1998gf}
\begin{equation}
    b^\mu \rightarrow b'^{\mu} = b^\mu + i s_2^\mu\,,
\end{equation}
and we invite the readers to see Refs.~\cite{Arkani-Hamed:2019ymq,Akhtar:2024mbg} for recent applications within the context of the KMOC formalism. This alters the angular momenta as 
\begin{align}
   &  l^\mu \rightarrow l'^\mu = \epsilon^{\mu \nu \rho \sigma}(b_{\nu} + i s_{2\nu})v_{1\rho}v_{2\sigma}\,, \\
   & l'\cdot\overline{l'} = (y^2-1)b^2 + (y^2-1)s_2^2 + (v_1\cdot s_2)^2\;.
\end{align}
Notice that the $l'\cdot\overline{l'}=-Q-2y\tilde{L}$, where an overline denotes complex conjugation, is a linear combination of $Q$ and $\tilde{L}$, which for a spinless probe read
%
\begin{align}
    Q\Big\rvert_{s_1 = 0} & = -(y^2-1)b^2 -2y(l \cdot s_2) - (y^2-1) s_2^2 - (v_1 \cdot s_2)^2\,, \\[0.1cm]
    \tilde{L}\Big\rvert_{s_1 = 0} & = (l\cdot s_2)\;.
\end{align}
Setting $s_1=0$ in \eqref{eq:asymptotic-L} gives $\tilde{L}=l\cdot s_2$, whose conservation follows from the axial symmetry of Kerr. Since $y=E/m_1$ is also conserved, the identity $l'\cdot\overline{l'}=-Q-2y\tilde{L}$ makes conservation of $l'\cdot\overline{l'}$ equivalent to conservation of the Carter constant for the scattering of a spinless probe in Kerr spacetime.

\section{1PM beyond probe: a new basis of conserved charges}
%
The conserved quantities in the main text are only in involution if we take the probe limit $m_2\rightarrow\infty$. This implies, for example, that although the tree-level radial action can be written in terms of the conserved quantities, integrability is not manifest outside of the probe limit. This begs the question: can we find a better basis of conserved quantities that are in involution beyond the probe limit? 
At least at 1PM the answer is yes. We find that the following three conserved quantities 
\begin{align}
    \mathscr{L}&=
    \frac{m_1^2 (l \cdot s_1)}{m_1^2+m_2^2} + \frac{m_2^2 (l \cdot s_2)}{m_1^2+m_2^2}-\frac{(2 m_1 m_2 +(m_1^2+m_2^2)y)}{m_1^2+m_2^2}(s_1\cdot s_2)+ (v_2 \cdot s_1)(v_1 \cdot s_2)\,,\\
    \mathscr{Q}_Y&=
    \frac{m_1^2 (l \cdot s_2)}{m_1^2+m_2^2}+\frac{m_2^2 (l \cdot s_1)}{m_1^2+m_2^2} +\frac{(2 m_1 m_2 +(m_1^2+m_2^2)y)}{m_1^2+m_2^2}(s_1\cdot s_2)- (v_2 \cdot s_1)(v_1 \cdot s_2)\,,\\
    \begin{split}
    \mathscr{Q}&=-l^2+\frac{2(m_1^2-m_2^2)y}{m_1^2+m_2^2} (l\cdot s_2)-(y^2-1)s_2^2-(v_1\cdot s_2)^2-\frac{2(m_1^2-m_2^2)y}{m_1^2+m_2^2} (l \cdot s_1 )\\
    &+2\left(y^2+1+\frac{4m_1 m_2 y}{m_1^2+m_2^2}\right)(s_1\cdot s_2)-2y(v_2 \cdot s_1)(v_1 \cdot s_2)-(y^2-1)s_1^2- (v_2 \cdot s_1)^2
    \end{split}
\end{align}
commute with the 1PM radial action to all orders in spin and are in involution for generic masses. In other words, for $F_i,F_j \in\{I_r^{(1)},\mathscr{L},\mathscr{Q}_Y,\mathscr{Q}\}$ we find $\{F_i,F_j\}_{\DB}=0$ without taking the probe limit. The quantities above are symmetric under swapping $1\leftrightarrow2$ and reduce to the usual $\tilde{L},Q_Y,Q$ when $m_2\rightarrow\infty$.
We can also rewrite the 1PM radial action in terms of these quantities as
\begin{equation}
\begin{split}
    I_{r}^{(1)}&=\frac{Gm_1m_2(1-2y^2-2y\sqrt{y^2-1})}{2\sqrt{y^2-1}}\log\left(\frac{2(y+\sqrt{y^2-1})}{y^2-1}\mathscr{L}-\frac{2(y-\sqrt{y^2-1})}{y^2-1}\mathscr{Q}_Y+\frac{1}{y^2-1}\mathscr{Q}\right)\\
    &+\frac{Gm_1m_2(1-2y^2+2y\sqrt{y^2-1})}{2\sqrt{y^2-1}}\log\left(\frac{2(y-\sqrt{y^2-1})}{y^2-1}\mathscr{L}-\frac{2(y+\sqrt{y^2-1})}{y^2-1}\mathscr{Q}_Y+\frac{1}{y^2-1}\mathscr{Q}\right)\,.
\end{split}
\end{equation}
At 2PM (or 3PM) and up to $\mathcal{O}(s^2)$ one can find three quantities that commute with the $\cO(G^2)$ (or $\cO(G^3)$) piece of the radial action beyond the probe limit. However, these quantities do not commute with the radial action at any other order in $G$, in particular they do not commute with the tree-level radial action. This then suggests that it would be possible, as for the case of the 2PN spinning Hamiltonian~\cite{Tanay:2020gfb}, to find some perturbative integrability for generic spinning binaries in the post-Minkowskian expansion.


\end{document}